\documentclass{ws-ijmpcs}

\usepackage{amsmath}
\usepackage{amsfonts}
\usepackage{graphicx}
\usepackage{slashed}
\usepackage{dcolumn}
\usepackage{bm}
\usepackage{amssymb}
\usepackage{latexsym}
\usepackage{color}
\usepackage{url}

\newcommand{\pbrac}[1]{\left( #1 \right)}
\newcommand{\tbrac}[1]{\left[ #1 \right]}
\newcommand{\cbrac}[1]{\left\{ #1 \right\}}

\newcommand{\bfrac}[2]{{\left(\frac{#1}{#2} \right)  }}

\begin{document}

\markboth{Arsham Farzinnia}
{Classically Scale Invariant Inflation and (A)gravity}

\catchline{}{}{}{}{}
%
%%%%%%%%%%%%%%%%%%%%%%%%%%%%%%%%%%%%%%%%%%%%%%%%%%%%%%%%%%%%%%%%%%%%

\title{Classically Scale Invariant Inflation and (A)gravity}

\author{Arsham Farzinnia\footnote{Speaker}}

\address{Center for Theoretical Physics of the Universe, Institute for Basic Science (IBS)\\Daejeon 305-811, Republic of Korea\\
farzinnia@ibs.re.kr}

\maketitle

\begin{history}
\received{Day Month Year}
\revised{Day Month Year}
\published{Day Month Year}
\end{history}

\begin{abstract}
In this talk, I present the minimal classically scale-invariant and $CP$-symmetric extension of the standard model, containing one additional complex gauge singlet and three flavors of right-handed Majorana neutrinos, incorporated within a renormalizable framework of gravity, consistent with these symmetries; the Agravity. I particularly focus on the slow-roll inflationary paradigm within this framework, by identifying the pseudo-Nambu-Goldstone boson of the (approximate) scale symmetry with the inflaton field, constructing its one-loop effective potential, computing the slow-roll parameters and the inflationary observables, and demonstrating the compatibility of the small field inflation scenario with the latest Planck collaboration data sets.\footnote{This contribution is based on the work in collaboration with Seyen Kouwn \cite{Farzinnia:2015fka}.}
\end{abstract}

\ccode{PACS numbers:}

\section{Introduction}

The classical scale invariance formalism has been considered as a (approximate) symmetry to protect various putative scales within Nature from a mutual quadratic destabilization, posing as an economical solution to the hierarchy problem \cite{Bardeen}. In this treatment, we consider a minimal extension of the standard model (SM) by adding a complex gauge singlet scalar and three flavors of the right-handed Majorana neutrinos, while imposing the $CP$ as well as the classical scale symmetry. The model\footnote{The TeV scale phenomenology of this minimal framework, without including gravity, were previously studied in \cite{Farzinnia:2013pga,Farzinnia:2014xia,Farzinnia:2015uma}, and the possibility for obtaining a strongly first-order electroweak phase transition, relevant for baryogenesis, was analyzed in \cite{Farzinnia:2014yqa}.} is embedded with the renormalizable framework of Agravity \cite{Salvio:2014soa}, which also respects the imposed symmetries. While the imposed $CP$ invariance prevents the singlet pseudoscalar from decaying, rendering it a stable dark matter candidate, nonzero masses for the SM neutrinos may be generated using the seesaw mechanism \cite{seesaw}. A nonzero singlet vacuum expectation value (VEV) is dynamically generated \textit{a la} Coleman-Weinberg \cite{Coleman:1973jx}, which subsequently induces the (reduced) Planck scale via the scalar non-minimal couplings, as well as the weak scale by means of the Higgs portal operators, accommodating the discovered 125~GeV state \cite{LHCnew}. It also generates appropriate mass terms for the dark matter, and the right-handed Majorana neutrinos. The scale symmetry protects the various scales from a mutual quadratic destabilization.

We explore the slow-roll inflationary paradigm within the introduced framework, by identifying the pseudo-Nambu-Goldstone boson of the (approximate) scale symmetry with the inflaton field, and construct the one-loop effective potential for this field. Trans-Planckian field excursions are accommodated within the Agravity framework. We compute the slow-roll parameters and the inflationary observables, and demonstrate the viability of the inflationary paradigm within the introduced model, in accordance with the latest observational data, published by the Planck collaboration \cite{Ade:2015lrj}.

\section{Formalism}

We minimally extend the SM content by introducing one complex gauge singlet scalar. The full scalar potential is conjectured to respect the $CP$ as well as the scale symmetry, and contains the following operators in its most general form
\begin{equation}\label{V0}
\begin{split}
V^{(0)}(H,S) =&\, \frac{\lambda_1}{6} \pbrac{H^\dagger H}^2 + \frac{\lambda_2}{6} |S|^4 + \lambda_3 \pbrac{H^\dagger H}|S|^2 + \frac{\lambda_4}{2} \pbrac{H^\dagger H}\pbrac{S^2 + S^{*2}} \\
&+ \frac{\lambda_5}{12} \pbrac{S^2 + S^{*2}} |S|^2 + \frac{\lambda_6}{12} \pbrac{S^4 + S^{*4}} \ ,
\end{split}
\end{equation}
where, $H$ and $S$ are the SM doublet and singlet, respectively,
\begin{equation}\label{HS}
H= \frac{1}{\sqrt{2}}
\begin{bmatrix} \sqrt{2}\,\pi^+ \\ \pbrac{v_\phi+\phi}+i\pi^0 \end{bmatrix} \ , \qquad S =\frac{1}{\sqrt 2} \tbrac{\pbrac{v_\eta + \eta} + i\chi} \ .
\end{equation}
The nonzero VEVs of the $CP$-even scalars are generated dynamically ($v_{\phi}=246$~GeV); in particular, once the singlet VEV, $v_{\eta}$, is induced via the Coleman-Weinberg mechanism \cite{Coleman:1973jx}, it is transmitted to the SM sector via the Higgs portal operators with the coefficients $\lambda_{3,4}$, leading to a mass term for the Higgs field, $\mu_H^{2}$, triggering the spontaneous breaking of the electroweak symmetry
\begin{equation}\label{muSM}
\mu_H^{2} = \frac{\lambda_{3}+\lambda_{4}}{2}\, v_{\eta}^{2} \equiv \frac{\lambda_{m}^{+}}{2}\, v_{\eta}^{2} \qquad \pbrac{\lambda_{m}^{+} < 0} \ .
\end{equation}

Interestingly, the $CP$ symmetry protects the pseudoscalar singlet, $\chi$, from decaying, rendering it a stable dark matter candidate \cite{Farzinnia:2013pga}. In addition, one may include three flavors of the right-handed Majorana neutrinos, $\mathcal{N}_{i} = \mathcal{N}_{i}^{c}$, in order to give masses to the SM neutrinos by means of the see-saw mechanism \cite{seesaw}. Assuming a $CP$- and scale-symmetric singlet sector, the right-handed Majorana neutrino masses may be generated via their Yukawa interactions with the singlet
\begin{equation}\label{LRHN}
\mathcal{L}_{\mathcal N} = \text{kin.} - \tbrac{Y^\nu_{ij}\, \bar{L}_{\ell}^{i} \tilde{H} \mathcal{N}^{j} + \text{h.c.}} -\frac{1}{2}y_{N} \pbrac{S + S^*} \bar{\mathcal{N}}^{i}\mathcal{N}^{i} \ ,
\end{equation}
where, for simplicity, a flavor-universal Yukawa coupling, $y_{N}$, is considered.

In order to consistently account for the gravitational contributions, pertinent to the inflationary paradigm and the trans-Planckian excursion of the inflaton, the described minimal extension of the SM is entirely embedded within the renormalizable, and $CP$- and scale-symmetric framework of the Agravity \cite{Salvio:2014soa}. In the Jordan frame, the full action is of the form
\begin{align}
\sqrt{|\det g|}\, \mathcal L^{J}= &\,\sqrt{|\det g|} \,\bigg\{ |D_\mu H|^2 + |\partial_\mu S|^2 - V^{(0)}(H,S) - \xi_H (H^\dagger H) R - \xi_S^1 |S|^2 R \notag \\
&- \frac{1}{2} \xi_S^2 (S^2+S^{*2}) R + \frac{R^2}{6 f_0^2} + \frac{\frac{1}{3}R^2 - R_{\mu\nu}^2}{f_2^2} + \mathcal{L}_{\mathcal N} + \mathcal{L}_{\text{SM}}^{\text{rest}} \bigg\} \ ,\label{LJ}
\end{align}
where, in addition to the previously introduced scalar potential, the Majorana neutrino and the remaining SM Lagrangians, one includes the non-minimal couplings of the scalars to the curvature, $\xi_H$ and $\xi_S^{1,2}$, as well as the pure Agravity (higher-derivative) operators with the dimensionless couplings $f_{0,2}^{2}$. While the $f_{0}^{2}$~operator gives rise to a scalar graviton, the $f_{2}^{2}$~operator (the Weyl term) produces the usual massless spin-2 graviton together with its ``Lee-Wick''~(LW) partner \cite{LW}, $\theta$, which is a massive spin-2 ghost with a negative norm. One notes that the Einstein-Hilbert term, $-\dfrac{\bar{M}_\text{P}^2}{2} R$, is absent in \eqref{LJ} due to the classical scale invariance requirement. However, the non-minimal couplings induce such a term, once the nonzero VEVs are obtained. Therefore, both the (reduced) Planck and the weak scales are induced dynamically via the VEV of the singlet within the current framework.

Performing the local Weyl transformation, one can show that the Einstein frame action, with the canonical Einstein-Hilbert term manifestly exhibited, reads
\begin{align}
&\sqrt{|\det g^E|}\, \mathcal L^E = \sqrt{|\det g^E|} \,\bigg\{ \mathcal L_\Phi^\text{kin} - V^{(0)E} + \frac{\frac{1}{3}(R^E)^{2} - (R_{\mu\nu}^{E})^{2}}{f_2^2} -\frac{\bar{M}_\text{P}^2}{2} \, R^E \notag\\
&\qquad\qquad\qquad\qquad\qquad\qquad + \mathcal{L}^E_{\mathcal N} + \mathcal{L}_{\text{SM}}^{\text{rest}, \,E} \bigg\} \ , \label{LE} \\
&\mathcal L_\Phi^\text{kin} \equiv \frac{6\bar{M}_\text{P}^2}{\zeta^2} \cbrac{|D_\mu H|^2 + |\partial_\mu S|^2 +\frac{1}{2} (\partial_\mu \zeta)^2} \ , \label{Lkinphi} \\
&V^{(0)E} \equiv\frac{36\bar{M}_\text{P}^4}{\zeta^4} \cbrac{ V^{(0)} + \frac{3}{8} f_0^2 \tbrac{\frac{\zeta^2}{6} - 2\xi_H (H^\dagger H) - 2\xi_S^1 |S|^2 - \xi_S^2 (S^2+S^{*2})}^2} \label{VE},
\end{align}
with $\zeta$~the scalar graviton in its ``conformal'' definition, and $v_{\zeta}^{2} = 6\bar{M}_\text{P}^2 $.

The physical VEV of the system can be pinpointed by employing the Gildner-Weinberg procedure \cite{Gildener:1976ih}, where one initially minimizes the tree-level potential \eqref{VE}. This defines the flat direction of the potential at a particular energy scale, $\Lambda$. The one-loop corrections then predominantly matter only along this flat direction, where they lift the flatness and specify the true vacuum. Performing the tree-level minimization, one obtains
\begin{equation}\label{lmpleta}
\lambda_m^+  (\Lambda)= -\frac{\lambda_\phi (\Lambda)}{3}\, \frac{v_\phi^2}{v_\eta^2} + \dots \ , \qquad
\lambda_\eta  (\Lambda)= \lambda_\phi (\Lambda)\, \frac{v_\phi^4}{v_\eta^4} + \dots \ , 
\end{equation}
where, the ellipses correspond to one-loop contributions due to fixing the cosmological constant to zero at one-loop. Since a classical scale-invariant treatment of the cosmological constant problem is outside the scope of this work, such terms should be simply neglected. The relations \eqref{lmpleta} are examples of the dimensional transmutation phenomenon, and demonstrate the absence of a quadratic sensitivity of various scales to one another. In particular, the weak scale is not quadratically sensitive to the singlet VEV scale, since the $\lambda_m^+$~mixing coupling is not a free parameter.

The three $CP$-even scalars of the framework, $\phi,\eta,\zeta$, are mixed with one another in the quadratic part of the Lagrangian \eqref{VE}, and can be rotated into the physical mass basis using an orthogonal $3\times3$~matrix with the mixing angles $\omega_{1,2,3}$,
\begin{equation}\label{scaldiag}
\begin{pmatrix} \phi\\ \eta\\ \zeta \end{pmatrix}
= \mathcal R\pbrac{\omega_{1},\omega_{2},\omega_{3}} \begin{pmatrix} h \\ \sigma \\ \kappa \end{pmatrix} \ ,
\end{equation}
where, $h$ corresponds to the physical 125~GeV state discovered by the LHC \cite{LHCnew}, $\sigma$ is massless at tree-level, serving as the pseudo-Nambu-Goldstone boson of the (approximate) scale symmetry, and $\kappa$ denotes the physical scalar graviton. One, subsequently, obtains along the flat direction (c.f \eqref{lmpleta})
\begin{equation}\label{t1t2}
\tan^{2}\omega_1 = \frac{v_\phi^{2}}{v_\eta^{2}} \ , \qquad \tan^{2}\omega_2 = \frac{v_\zeta^{2}}{v_\phi^2 +v_\eta^2} \ ,
\end{equation}
once more, signifying the absence of a quadratic destabilization among the separate dynamically-induced scales.

The pseudo-Nambu-Goldstone boson of the (approximate) scale symmetry, $\sigma$, can be expressed in terms of the radial combination of the field basis scalars
\begin{equation}\label{sigma}
\sigma^{2} = \phi^{2} + \eta^{2} + \zeta^{2} \ .
\end{equation}
The kinetic term of this scalar has a non-canonical form, $\frac{6\bar{M}_\text{P}^2}{\zeta^2} \tbrac{\frac{1}{2} (\partial_\mu \sigma)^2} $ (c.f. \eqref{Lkinphi}); however, using the field redefinition (c.f. \eqref{t1t2} and \eqref{sigma})
\begin{equation}\label{sc}
\sigma_{c} - v_{\sigma_{c}} = \int_{v_{\sigma}}^{\sigma} \frac{\sqrt6 \bar{M}_\text{P}}{\sin\omega_2} \frac{d\sigma'}{\sigma'} = \frac{\sqrt6 \bar{M}_\text{P}}{\sin\omega_2} \log \frac{\sigma}{v_\sigma} \ ,
\end{equation}
its kinetic term can be brought into the canonical form, $\frac{1}{2} (\partial_\mu \sigma_{c})^2$. Note that the kinetic term of the $\sigma$~boson becomes canonical at the minimum; hence, one can set $v_{\sigma_{c}} = v_{\sigma} = \frac{\sqrt6 \bar{M}_\text{P}}{\sin\omega_2}$.

Along the flat direction, the one-loop contributions to the potential can be written according to
\begin{align}
V^{(1)}(\sigma_{c}) =&\, \frac{\mathcal M^{4}}{64\pi^{2}\, v_{\sigma}^{4}} \, \sigma_{c}^4 \tbrac{\log \frac{\sigma_{c}^2}{v_{\sigma}^2} - \frac{1}{2}} \ , \label{V1fin} \\
\mathcal M^{4}\equiv&\, 5 M_{\theta}^{4}+ M_{\kappa}^{4}+ M_{\chi}^{4} -6 M_{\mathcal N}^{4}+ M_{h}^{4}+6M_{W}^{4} +3M_{Z}^{4} -12M_{t}^{4} \label{Bmodel} \ ,
\end{align}
where, the heavy SM degrees of freedom in the loop are also taken into account for completeness. The one-loop corrections \eqref{V1fin} is negative at its minimum; however, the tree-level potential \eqref{VE} is non-vanishing along the flat direction, due to the gravitational contributions, and can be utilized to fix the minimum of the full one-loop potential (the cosmological constant at one-loop) to zero. Therefore, one obtains for the full one-loop effective potential along the flat direction
\begin{equation}\label{V01}
V(\sigma_{c}) = \frac{\mathcal M^{4}}{128\pi^{2}} \tbrac{ \frac{\sin^{4}\omega_{2}}{36\, \bar{M}_\text{P}^4} \, \sigma_{c}^4 \pbrac{2\log \frac{\sigma_{c}^2\, \sin^{2}\omega_{2}}{6\bar{M}_\text{P}^2} - 1}+1} \ .
\end{equation}
This potential induces a radiative mass term for the $\sigma$~scalar
\begin{equation}\label{msigfin}
m_{\sigma}^{2} = \frac{\sin^2\omega_2}{48\pi^2} \, \frac{\mathcal M^{4}}{\bar{M}_\text{P}^2}  \ ,
\end{equation}
and is bounded from below (and positive definite) for $\mathcal M^{4} >0$, implying
\begin{equation}\label{massrel}
5M_{\theta}^{4} +M_\kappa^4 + M_\chi^4 - 6M_{\mathcal N}^4 >  12M_t^4- M_h^4 - 6M_W^4 - 3M_Z^4 \simeq (300~\text{GeV})^{4}\ ,
\end{equation}
which, also yields a non-tachyonic mass for the $\sigma$~boson.

\section{Inflation}

The pseudo-Nambu-Goldstone boson of the (approximate) scale symmetry in its canonical form, $\sigma_{c}$, may be identified with the inflaton of the framework, possessing the one-loop effective potential \eqref{V01} along the flat direction. From this potential the slow-roll parameters can be easily computed
\begin{equation}\label{slowpa}
\epsilon \equiv \frac{\bar{M}_\text{P}^2}{2}\bfrac{V_{\sigma_c}}{V}^2  \ , \qquad
\eta \equiv \bar{M}_\text{P}^2 \, \frac{V_{\sigma_c\sigma_c}}{V} \ , \qquad
\xi^2 \equiv \bar{M}_\text{P}^4 \, {V_{\sigma_c} V_{\sigma_c\sigma_c\sigma_c}\over V^2} \ , 
\end{equation}
where, the field subscripts denote taking the appropriate derivative(s) of the potential with respect to the argument.

Additionally, the number of $e$-foldings is easily found using the slow-roll equations of motion, and is given by
\begin{equation}\label{efdN}
\begin{split}
N \simeq &\, \frac1{\bar{M}_\text{P}^2} \int_{\sigma_{c,e}}^{\sigma_{c,i}}
\frac{V}{V_{\sigma_c} } \,
d{\sigma_c} \\
&= \cbrac{
\frac{3}{8 \sin^2 \omega_2}
\left[
{\rm Ei}\left(-\log \frac{\sin^2 \omega_2\,\sigma_{c}^2}{6 \bar{M}_\text{P}^2}\right)
-{\rm li}\left(\frac{\sin^2 \omega_2\,\sigma_{c}^2}{6 \bar{M}_\text{P}^2}\right)
\right]
+\frac{\sigma_{c}^2}{8 \bar{M}_\text{P}^2}} \Bigg|_{\sigma_{c,e}}^{\sigma_{c,i}} \ ,
\end{split}
\end{equation}
with $\sigma_{c,i}$ and $\sigma_{c,e}$ the values of the $\sigma_{c}$~inflaton at the beginning and
at the end of the inflation, respectively, ${\rm Ei}(z)$ the exponential integral (${\rm Ei}(z)=-\int_{-z}^{\infty}{e^{-t}\over t}dt$), and ${\rm li}(z)$ the logarithmic integral (${\rm li}(z)=\int_{0}^{z}\frac{1}{\log t} dt$). The inflation comes to an end, once the condition $\epsilon|_{\sigma_c=\sigma_{c,e}} \sim 1$ is reached. The inflaton field value at the horizon crossing point, $\sigma_c=\sigma_{c,i}$, can subsequently be obtained from \eqref{efdN} for a given $e$-folding number, as a function of the mixing angle $\omega_{2}$.

The amplitude of the scalar perturbations, $A_s$,
the scalar spectral index, $n_s$, 
the tensor-to-scalar ratio, $r$, 
and its running, $\alpha_s = d n_s / d \log k$,
are defined according to
\begin{equation}\label{inflobs}
A_s =\frac{V_*}{24\pi^2\bar{M}_\text{P}^4 \,  \epsilon_*} \quad n_s-1 = -6\epsilon_*+2\eta_* \quad
r =16\epsilon_*  \quad \alpha_s = 16 \epsilon_* \eta_* -24 \epsilon_*^2 -2 \xi_*^2 \ ,
\end{equation}
with the subscripted asterisk denoting the field value $\sigma_c=\sigma_{c,i}$.
Given the inflaton potential \eqref{V01} and the slow-roll parameter definitions \eqref{slowpa}, these quantities are easily determined, as functions of $\sin \omega_2$, $\mathcal M$, and the field value $\sigma_{c,i}$ at the horizon exit.

\begin{figure}
\centerline{\includegraphics[width=.8\textwidth]{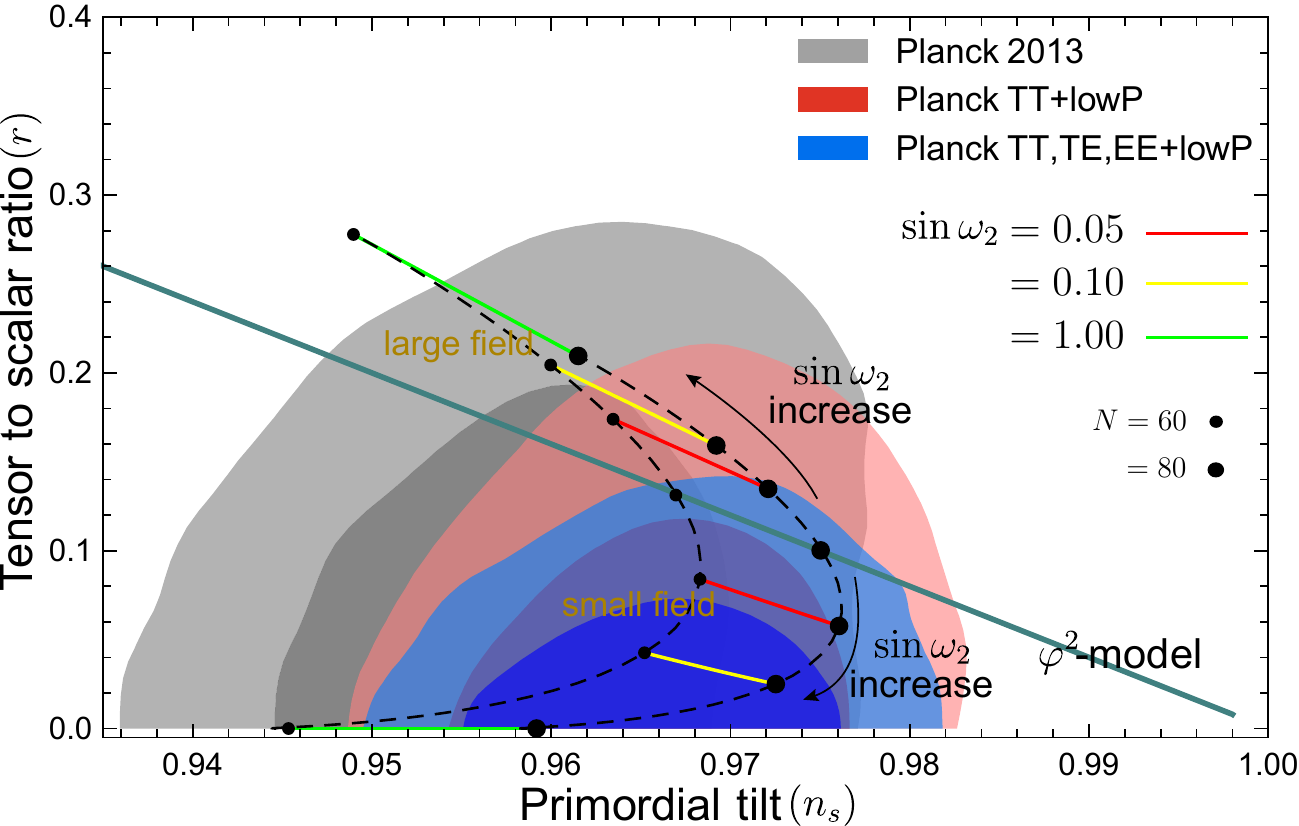}}
\vspace*{8pt}
\caption{Several observational data sets by the Planck collaboration at 68\%~C.L. (darker region) and at 95\%~C.L. (lighter region) plotted within the $n_s - r$~plane, along with the predictions of the current framework for both the small and large field inflation scenarios (dashed lines). Two $e$-folding number benchmarks, $N=60,80$, are selected and the full range of the mixing angle, $\sin\omega_{2}$, is considered. The diagonal line labeled as ``$\varphi^{2}$-model'' denotes the chaotic inflation scenario, in the $\sin\omega_{2} \to 0$ limit. \label{fig:nsr}}
\end{figure}

The predictions of the current framework for these observables are displayed within the $n_s - r$~plane of Fig.~\ref{fig:nsr}, for the small and the large field inflation scenarios. Several benchmark values of the mixing angle are exhibited for $N=60,80$. In addition, the observational constraints from several Planck collaboration published data sets at 68\% and 95\%~C.L. \cite{Ade:2015lrj} are incorporated within the plot. In the limit $\sin\omega_{2} \to 0$, the inflaton potential \eqref{V01} reduces to the ordinary chaotic inflation scenario near its minimum
\begin{equation}
V(\sigma_c) \simeq 
{{\mathcal M}^4 \sin^2 \omega_2 \over 96 \pi^2 \bar{M}_\text{P}^2}
\left(\sigma_c-{\sqrt{6} \bar{M}_\text{P} \over \sin \omega_2}\right)^2 \ ,
\end{equation}
with the limiting values
\begin{equation} 
(n_s, r)\Big |_{\sin\omega_{2}\to 0} \simeq \left(1-{4\over 1+ 2N}, {16\over 1+ 2N}\right) \ .
\end{equation}
Therefore, we have $(n_s,r) = (0.967,0.132)$ for $N=60$, and $(n_s,r) = (0.975,0.099)$ for $N=80$. In contrast, the $\sin\omega_{2} \to 1$~limit is for the most part compatible with the observational data within the small field inflation scenario. While in the large field inflation scenario the potential becomes increasingly steeper, the small field inflation scenario is characterized by a relatively flat potential with a very small $\epsilon$~slow-roll parameter. Hence the $\eta$~slow-roll parameter dominates the behavior of the spectral index (c.f.~\eqref{inflobs}), and one obtains for the leading-order behavior in this scenario
\begin{equation} 
(n_s, r)\Big |_{\sin\omega_{2}\to 1} \simeq \left(1-{3\over N}, 0\right) \ .
\end{equation}

\section{Conclusion}

A minimal extension of the SM has been developed by introducing one complex singlet scalar and three (mass-degenerate) right-handed Majorana neutrinos. The singlet sector is postulated to respect the $CP$ as well as the classical scale symmetry, and is embedded within the renormalizable Agravity framework. The $CP$ symmetry prohibits a decay of the singlet pseudoscalar, rendering it a stable dark matter candidate, whereas the SM neutrinos obtain nonzero masses via the seesaw mechanism.

The singlet VEV, dynamically generated via the Coleman-Weinberg mechanism, simultaneously induces the weak scale (via the Higgs portal operators) as well as the (reduced) Planck scale (via the scalar non-minimal couplings to the curvature). Furthermore, it generates mass terms for the right-handed Majorana neutrinos (via Yukawa interactions) and the dark matter. It is shown that the scale symmetry protects the various scales from a mutual quadratic destabilization. 

Identifying the pseudo-Nambu-Goldstone boson of the (approximate) scale symmetry of the model, in its canonical form, with the inflaton field, its one-loop effective potential is constructed, and the viability of the slow-roll inflationary paradigm is validated; in particular, a trans-Planckian field excursion is consistently accommodated utilizing the Agravity framework. It is demonstrated that small field inflation scenario can be rendered fully compatible with the latest observational data for the suitable $e$-folding numbers, within the minimal classically scale invariant framework.

%\begin{thebibliography}{000} %for 3 digits
%\begin{thebibliography}{00}  %for 2 digits

\end{document}